\documentclass[pdflatex,sn-nature]{sn-jnl}

\usepackage{graphicx}%
\usepackage{multirow}%
\usepackage{amsmath,amssymb,amsfonts}%
\usepackage{amsthm}%
\usepackage{mathrsfs}%
\usepackage[title]{appendix}%
\usepackage{xcolor}%
\usepackage{textcomp}%
\usepackage{manyfoot}%
\usepackage{booktabs}%
\usepackage{algorithm}%
\usepackage{algorithmicx}%
\usepackage{algpseudocode}%
\usepackage{listings}%

\usepackage{soul}
\usepackage{titlesec}

\titleformat{\paragraph}[runin]
  {\normalfont\bfseries}{}{0em}{}
\titlespacing{\paragraph}{0pt}{\baselineskip}{0.5em}
\usepackage{xcolor}

\newif\ifdraft

\drafttrue      

\ifdraft
  \newcommand{\todo}[1]{\textcolor{red}{[TODO: #1]}}
  \newcommand{\note}[1]{\textcolor{blue}{[Note: #1]}}
  \newcommand{\ag}[1]{\textcolor{purple}{[Arnab: #1]}}
  
\else
  \newcommand{\todo}[1]{}
  \newcommand{\note}[1]{}
  \newcommand{\ag}[1]{}
\fi

\theoremstyle{thmstyleone}%
\theoremstyle{thmstyletwo}%

\theoremstyle{thmstylethree}%

\begin{document}

\title[Moire Spectrometer]{Broadband Content-Adaptive Moiré Meta-spectrometer}


\author[1]{\fnm{Arnab} \sur{Ghosh}}\email{aghos034@ucr.edu}
\author[2]{\fnm{Johannes E.}\sur{Fr\"och}\email{jfroech@uw.edu}}
\author[2]{\fnm{Arka} \sur{Majumdar}}\email{arka@ece.uw.edu}

\author*[1]{\fnm{Vishwanath} \sur{Saragadam}}\email{vishwans@ucr.edu}

\affil*[1]{\orgdiv{Department of Electrical and Computer Engineering}, \orgname{University of California, Riverside}, \orgaddress{\street{900 University Ave}, \city{Riverside}, \postcode{92521}, \state{California}, \country{USA}}}

\affil[2]{\orgdiv{Department of Electrical and Computer Engineering}, \orgname{University of Washington}, \orgaddress{\street{185 Stevens Way}, \city{Seattle}, \postcode{98195}, \state{Washington}, \country{USA}}}






\maketitle

\begin{abstract}

Optical spectroscopy underpins material characterization, chemical sensing, and astronomy, but conventional instruments face a rigid trade-off between footprint, spectral range, and resolution. We demonstrate a content-adaptive spectrometer that overcomes this by co-designing dispersive Moiré meta-optics with a recursive sampling algorithm. Instead of using Moiré metalenses solely for varifocal tuning, we harness the strong chromatic aberration arising from phase-wrapping in their subwavelength metasurface architecture. This hyperchromaticity enables a deterministic, one-to-one mapping between the metasurfaces’ mutual rotation angle and the sharply focused wavelength, repurposing the pair as a high-resolution spectral scanner. To accelerate data acquisition, we introduce a content-adaptive recursive sampling protocol that exploits the structural sparsity of physical spectra: a fast coarse sweep identifies high-information regions, followed by successively finer angular refinement only where needed. Using a laboratory prototype spanning 405–980\,nm, we reconstruct diverse spectra—from smooth broadband to sparse multi-line laser emissions—with nearly 3× fewer measurements on average at matched fidelity (up to 7× for sparse line spectra), achieving 30 dB reconstruction 6.7× faster than conventional uniform sampling. This establishes a framework for intelligent, task-adaptive meta-optical sensors that tightly integrate physical dispersion with computational signal processing for real-time spectrometry.

\end{abstract}

\section {Introduction}\label{intro}

\begin{figure}[!tt]
    \centering
    \includegraphics[width=1\linewidth]{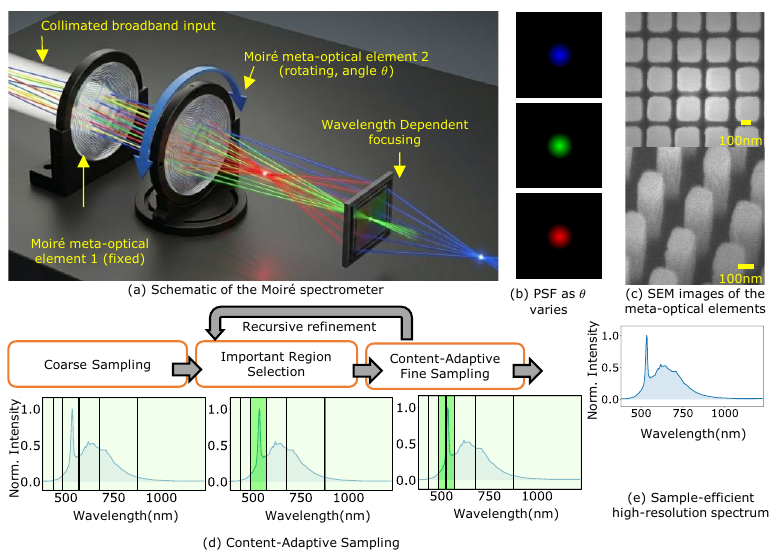}
    \caption{\textbf{A content-adaptive Moir\'e Meta-optical spectrometer}. A pair of rotating Moir\'e optics enable focus tuning; by exploiting the chromatic properties of meta-optics, we can instead tune the specific wavelength that is in focus, thereby enabling a fast and flexible single pixel spectrometer. \textbf{a}) Optical layout of the rotational spectrometer showing the rotating meta-optic (MO1) cascaded with a static meta-optic (MO2) at close proximity to each other. The relative rotation of MO1 with respect to MO2 causes wavelength dependent focus shift at the sensor plane. \textbf{b}) PSF changes from blue to infrared wavelengths as the angle between the two elements increases. \textbf{c}) SEM images of the meta-optical elements showing the nano-pillars. \textbf{d)} shows our content adaptive sensing strategy, where we first perform coarse sampling, which enables estimation of important regions (such as laser peak). We then scan only within these important regions, and repeat the process recursively. \textbf{e)} The end product is a high resolution spectrum with significantly fewer measurements compared to full Nyquist sampling. }
    \label{fig:1-theory_and_system_design}
\end{figure}

 Broadband compact spectrometers that combine small form factor with high spectral resolution are increasingly gaining popularity in the research community and show a lot of promise in wearable portable sensing, diagnostics and field deployable instrumentation \cite{tran_compact_2023}. Conventional dispersive spectrometers use diffraction gratings or prisms to separate wavelengths and sample the dispersed light uniformly across the band, which constrains both spectral resolution and optical throughput, increases the overall form factor, and prevents the instrument from distinguishing wavelength regions that carry a lot of information from those that carry relatively little~\cite{ye_miniature_2016}. Recent advances in meta-optics have paved the way for ultra-thin, multifunctional spectral devices~\cite{yu_light_2011}: metalenses offer a fundamentally higher degree of design freedom over phase, wavelength, and chromatic dispersion than classical diffractive optical elements (DOEs)~\cite{chen_broadband_2018,khorasaninejad_achromatic_2017}, but the sampling axis of these meta-optical instruments is still fixed at fabrication, and compact spectrometers therefore continue to balance spectral range, spectral resolution, and acquisition time
against the non-uniform information content of the signals they measure.

Content-aware sampling is a natural solution to this trade-off between resolution and time. The instrument can allocate fewer samples to smooth regions and more samples to sharper regions that need refinement, so that the overall sample budget is concentrated where the spectrum is informative. Wavelet-tree adaptive sampling has been demonstrated for hyperspectral images, ghost imaging, and free-form lensing~\cite{dai_adaptive_2014,yu_adaptive_2014,saragadam_wavelet_2019}, compressed hyperspectral acquisition exploits the same structural sparsity~\cite{hahn_compressive_2014}, active wavelength selection has been shown in tunable-laser chemical sensing~\cite{sullivan_active_2015}, autonomous planetary spectroscopy on Mars uses adaptive acquisition to manage limited measurement budgets in the field~\cite{lawson_adaptive_2025}, and Gaussian-process-driven autonomous experiments are now routine at large-scale beamline facilities~\cite{noack_gaussian_2021,teixeira-parente_active_2023}. A small fraction of the spectral axis carries most of the information across all of these settings, and that is the structural property our scanner exploits.

Our key insight is that this content aware programmability can be realized by combining the Moir\'e
varifocal effect~\cite{bernet_adjustable_2008,ogawa_rotational_nodate} with the adaptive scanning algorithm. The relative rotation between two cascaded meta-optical elements allows us to move on the spectral axis independent of the lens geometry and select the wavelength that will be in focus in the sensor during acquisition time, and the motor step size sets the spectral sampling interval. The adaptive scanning algorithm then exploits this axis directly, choosing where the next measurement is taken based on the spectrum observed so far~\cite{saragadam_wavelet_2019,sullivan_active_2015}. The two are tightly coupled in a closed acquisition loop, with the algorithm dynamically reprogramming the dispersion law in response to the measured spectrum. The result is a compact, single-pixel meta-optical spectrometer~\cite{edgar_principles_2019} that allocates measurements where they matter most: smooth regions are captured with very few samples, sharp spectral features trigger localized refinement, and a usable spectrum is available after every acquisition round.
  
We realize the Moir\'e varifocal mechanism by cascading two meta-surfaces with equal-magnitude, opposite-signed quadratic-azimuthal phase profiles, mounted along a common optical axis. 
A relative rotation between the two metasurfaces synthesizes a net quadratic phase via the Moir\'e effect, producing an effective varifocal lens whose optical power scales linearly with the rotation angle. The rotation angle can then be mapped to the wavelength brought into focus at a fixed detector plane, based on the reciprocal relationship between focal length and wavelength due to wrapping of the Moir\'e optics phase profile~\cite{arbabi_controlling_2017}. The mapping is linear in wavenumber, so that the motor step size, not the geometry of the optic, sets the spectral sampling interval. A content-adaptive Faber-Schauder scanner then drives this substrate in closed loop: a content-independent pilot sweep partitions the band into intervals, leaf intervals are scored by a local second-difference detail coefficient, those whose detail exceeds an intensity-normalised threshold are bisected by acquiring a single new measurement at the midpoint, and the policy directly commands the rotation stage so that the dispersion law is reprogrammed on-the-fly in response to the spectrum being measured.

\section {Results }


The core enabling hardware for our adaptive scanner is a pair of Metasurfaces with Moir\'e phase functions that together implement a tunable spectral focusing element. We provide an overview of our hardware setup and then explain the content-adaptive sampling strategy.

\subsection{Moir\'e meta-optical spectrometer}
The spectrometer comprises two cascaded metalenses with equal and opposite phase profiles
$\Phi_{\mathrm{MO}_1}(r,\varphi) = a\,r^{2}\varphi$ and
$\Phi_{\mathrm{MO}_2}(r,\varphi) = -\Phi_{\mathrm{MO}_1}(r,\varphi)$~\cite{bernet_adjustable_2008}, mounted along a common optical axis
(Fig.~\ref{fig:1-theory_and_system_design}a). As $\mathrm{MO}_1$ is rotated with respect to the stationary $\mathrm{MO}_2$, the Moir\'{e} effect produces a quadratic phase whose optical power (inverse of focal length) scales linearly with the relative rotation angle~$\theta$~\cite{bernet_adjustable_2008,bernet_demonstration_2013} for a fixed wavelength.
Conversely, for a fixed focal length, the wavenumber ($1/\lambda$) changes linearly with the relative rotation angle.
We leverage this $\lambda-\theta$ relationship to implement a flexible spectrometer.
%
Our Meta-optical spectrometer is governed by the following wavelength-angle relationship(derived in Supplementary Note~1),
\begin{equation}
  \theta
  = \frac{\pi}{a\,f}\;\frac{1}{\lambda} + \theta_{0},
  \label{eq:calibration}
\end{equation}
where $f$ is the effective focal length and $\theta_0$ a constant angular offset. 
%
%
We calibrated our optical system with 6 narrowband lasers to obtain the following empirical relationship,
\begin{equation}
  \lambda\;[\mathrm{nm}]
  = \frac{10^{7}}{358.46\;\theta_{\mathrm{peak}}
    + 11\,761.70\,}.
  \label{eq:angle_to_wavelength}
\end{equation}
Given this one-to-one and continuous relationship, spectral resolution becomes a programmable parameter: coarser angular steps trade resolution for speed, while finer steps increase sampling density
without hardware modification. This programmability enables content-adaptive acquisition via a recursive sampling strategy, where an initial coarse sweep identifies spectrally rich regions and subsequent rounds allocate measurements preferentially to those regions
(Fig.~\ref{fig:1-theory_and_system_design}b--c). In contrast, a fixed resolution scanning either results in a highly coarse spectrum, missing out on important spectral details, or a very slow scan that requires substantially more samples. 

\begin{figure}[!tt]
    \centering
    \includegraphics[width=1\linewidth]{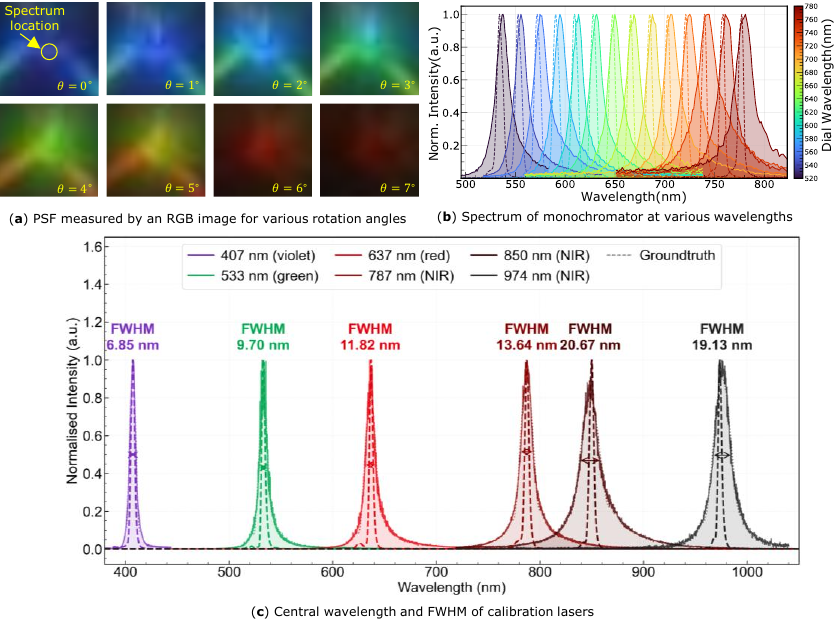}
    \caption{\textbf{Characterization of our Moir\'e Spectrometer.} (a) shows the PSF captured by a color camera at some representative angles between the two meta-optic. The circle in the top left image shows the location where the PSF had least rotation, and hence corresponded to the best location for sampling the spectrum. We note that all our measurements were captured with a grayscale camera; an RGB camera was used only to show the PSFs. (b) Monochromator swept at $14$ dial settings ($520$--$780$\,nm, $20$\,nm steps), measured simultaneously by our spectrometer (solid with fill) and an off-the-shelf reference(dashed); color encodes the dial wavelength. (c) shows FWHM and central wavelength of spectra of the calibration laser sources. As expected, the FWHM increases as we move from 400\,nm to 1000\,nm, as the resolution is constant in wavenumber. Across the board, we establish our optical setup as a precise spectrometer.}
    \label{fig:characterization}
\end{figure}


\subsection{Characterizing the Moir\'e spectrometer}
We calibrated the $\lambda-\theta$ relationship by scanning 6 narrowband lasers from 405\,nm to 980\,nm, enabling a visible to near infrared (NIR) spectrometer.
Figure~\ref{fig:characterization} (a) directly shows the PSF at some angles captured by an RGB camera for reference to demonstrate the effect of chromatic shift at fixed focal length. The center of the PSF shows a sharp change in color with angle in a range from $0^{\,\circ}$ to $7^{\,\circ}$, demonstrating the distinct $\lambda-\theta$ relationship.
Figure~\ref{fig:characterization} (b) shows the spectral accuracy of the system against an off-the-shelf reference spectrometer over $14$ monochromator dial settings from $520$--$780$\,nm. The dial settings are distinguished by color, with the system and ground truth denoted by solid and dashed lines, respectively. We observe that the system’s peak wavelength aligns with the reference peak value, apart from a constant offset of 3.6 nm. 
Figure~\ref{fig:characterization} (c) shows the FWHM and central wavelength of various calibration sources throughout the visible and NIR range to demonstrate the accuracy in spectral reconstruction, emphasizing the precision and resolution of our optical setup.
\begin{figure*}[!tt]
  \centering
  \includegraphics[width=1.0\linewidth]{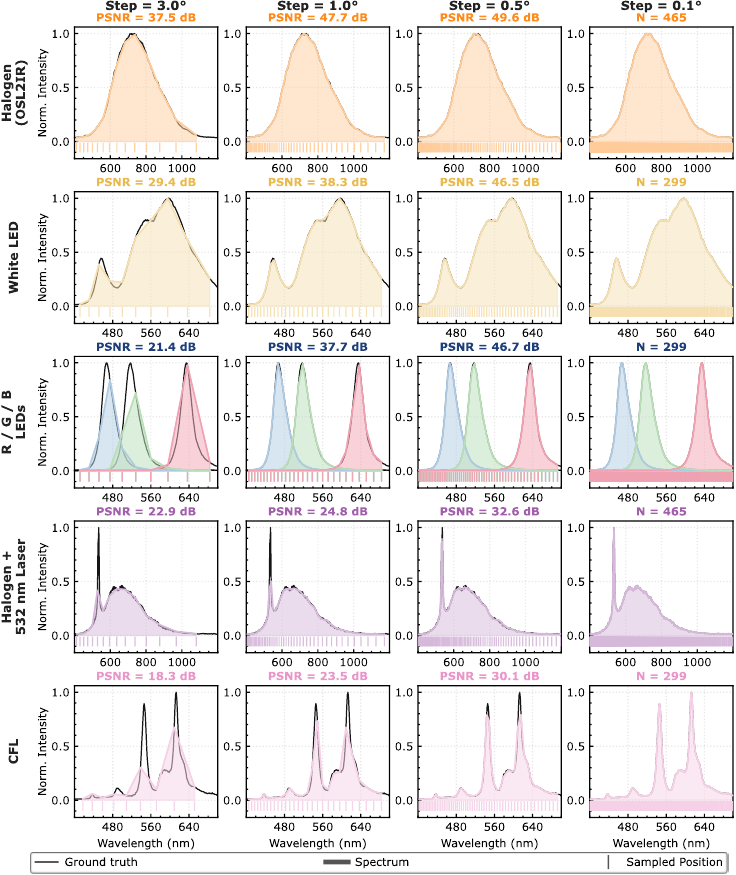}
  \caption{\textbf{Flexible-resolution spectrometry.} Reconstructed spectra of five light sources (rows: halogen, white LED, R/G/B LED triplet, halogen + 532\,nm laser, CFL) at angular step sizes $\Delta\theta = 3^{\circ}, 1^{\circ}, 0.5^{\circ}, 0.1^{\circ}$ (columns). Coloured curves are piecewise-linear reconstructions $\hat{f}(\lambda)$, row-normalized to the peak of the densest $0.1^{\circ}$ scan; the $0.1^{\circ}$ scan is overlaid on coarser-step panels in black as the in-figure ground truth. Ticks below each panel mark sampled angular positions; per-panel labels give the step size, PSNR against the $0.1^{\circ}$ reference (coarser columns), and $0.1^{\circ}$ sample count $N$ (rightmost column).}
  \label{fig:flexible-resolution}
\end{figure*}

\begin{figure}[!tt]
    \centering
    \includegraphics[width=\linewidth]{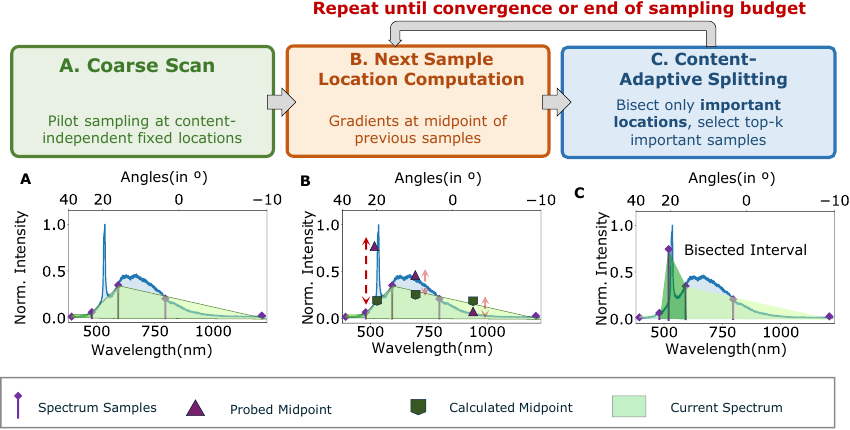}
    \caption{\textbf{Content-adaptive sampling.} We propose a two-stage, three step content-adaptive sampling strategy that enables high resolution and sample efficient scanning. A) We first do a content-independent pilot sample at very coarse resolution. B) Then, we estimate the next locations to sample by looking at local gradients of the spectrum. C) The spectrometer is then directed to sample only at these important locations. (B, C) are repeated until the spectrum converges, or the sampling budget ends. This enables an ``anytime" spectrometer where each round of content-adaptive sampling results in a usable spectrum.}
    \label{fig:flowchart}
\end{figure}

\begin{figure}[!tt]
    \centering
    \includegraphics[width=1\linewidth]{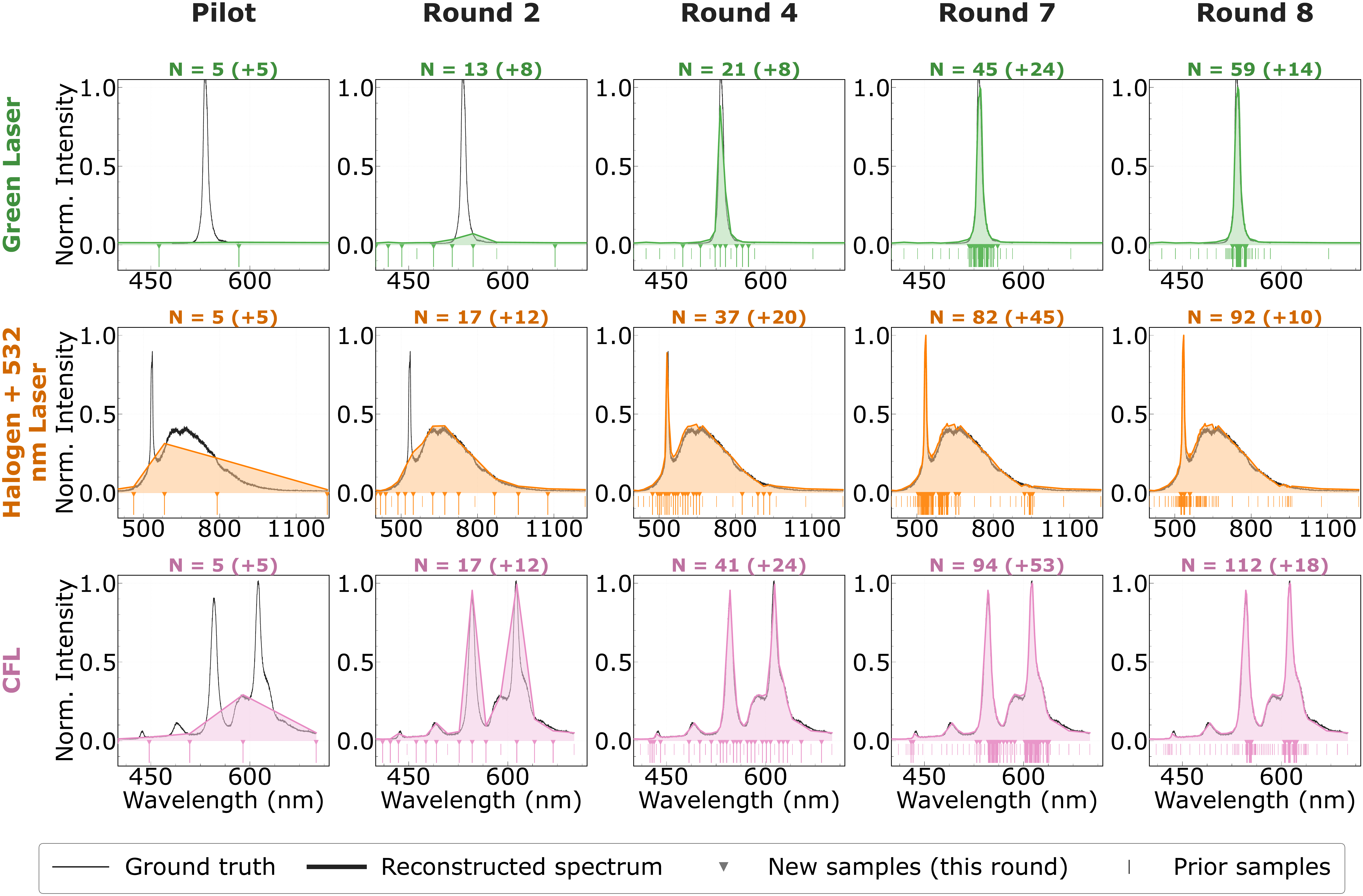}
    \caption{\noindent\textbf{Adaptive Faber-Schauder bisection algorithm based acquisition across three classes of test spectra.} Rows: a 532\,nm diode laser (top), a halogen continuum combined with a 532\,nm laser line (middle), and a compact fluorescent lamp (CFL, bottom). Columns show the reconstruction at the uniform pilot sweep and at three successive adaptive rounds; each panel reports the cumulative sample count $N$ and, in parentheses, the new samples added in that round. Filled curves are the piecewise-linear reconstruction $\hat{f}_t(\lambda)$ from the cumulative sample set. Triangles beneath the wavelength axis mark sample locations (long dash : current round; small dash : prior rounds). }
    \label{fig:adaptive_cfl}
\end{figure}

\subsection{Flexible resolution spectrometer}\label{sec:flexible-resolution}
A central feature of the platform is that spectral resolution functions as a programmable acquisition setting: by adjusting the motor step size, the setup can trade sampling density (and thus effective resolution) against measurement speed. We illustrate this by measuring the same spectral target with several different angular step sizes. 
Figure~\ref{fig:flexible-resolution} shows scans of various commonly occurring light sources at various angular steps, from very coarse $3^\circ$ that requires as few as 11 measurements, to very fine $0.1^\circ$ that requires as many as 301 measurements.
We first observe that our optical setup can scan any spectrum with high precision across a very broad range of wavelengths.
In particular, compact fluorescent lamps (CFLs) have strong emission peaks that are precisely identified by our optical setup.
Second, we make a critical observation about the trade-off between the resolution and number of samples.
When spectrum is smooth (like halogen) and spread across the whole wavelength range, the quality does not improve past a small set of measurements.
In contrast, for sharp spectra, quality keeps improving until $0.1^\circ$ step size, but the spectrum itself occupies an extremely narrow part of the wavelength range.
This combination is captured in the ``Halogen+532\,nm" source where the laser is refined as resolution progresses, but the halogen part does not improve.
This motivates our content-adaptive scanning strategy which we discuss next.

A useful consequence of the mapping between rotation and wavelength is that a uniform angular sweep automatically generates a wavelength-dependent resolution, fine at the blue end of the band and coarser at the red end. Because Eq.~\eqref{eq:calibration} is linear in wavenumber, equal angular steps result in a wavelength resolution that scales as $\Delta\lambda \propto \lambda^{2}$ (Supplementary Note~1). This is well-matched to typical spectroscopic content, in which atomic and ionic emission lines cluster more densely at short wavelengths where the same angular step delivers finer $\Delta\lambda$, while smooth continua are adequately sampled by coarser steps at long wavelengths. The content-adaptive scanning algorithm functions on top of this baseline by redirecting samples to regions of large spectral detail.

\subsection{Content-adaptive spectrometry}\label{subsec:content-adaptive-spectrometry}
Motivated by the need for varying sampling densities for smooth spectra and laser-like spectra, we develop a wavelet tree-based content-adaptive scanner.
This adaptive scanning follows two stages.
In the first stage, we capture a small set of uniformly (in wavenumber) spaced samples across the whole wavelength range to estimate the important regions.
These important regions are estimated by the ``detail coefficients" of the wavelet zero-tree~\cite{saragadam_wavelet_2019}, which is similar in principle to gradients in spectrum.
In the second stage, we measure samples only in the important regions to refine the spectrum.
This is primarily enabled by our fast and flexible scanning hardware that is otherwise not possible with a dispersive element-based spectrometer.
The second stage is performed recursively, where we re-estimate the important regions and then scan only there.
Figure~\ref{fig:flowchart} shows a flowchart of the proposed content-adaptive sampling strategy.
The full scanner and its parameter settings are detailed in Supplementary Note~2(Supplementary Algorithm~1 and Supplementary Table~1).
We show results of adaptively sampling common light sources in Fig.~\ref{fig:adaptive_cfl} across 8 rounds of adaptive sampling.
For smooth spectra such as halogen, the second stage terminates early, as the improvements are marginal.
For sharp laser-like spectra, the stages continue, but the number of samples per stage are considerably small as the region of sampling is compact.
Hence, adaptive sampling captures both the sharp laser peak and the smooth halogen continuum with roughly 3× fewer measurements on average---up to 7× for sparse line spectra---than uniform sampling at matched fidelity.
Supplementary Figs.~1--3 show the complete round-by-round reconstructions for a broadband-plus-laser source, a compact fluorescent lamp, and a 532~nm laser, while Supplementary Figs.~4--5 report the eleven-source adaptive-versus-uniform benchmark (Supplementary Note~3).
%


\section{Discussion}\label{sec:discussion}

We introduced a content-adaptive Moiré meta-spectrometer in which the spectral sampling axis is reshaped during acquisition by the algorithm, rather than being predetermined by the optics. Recent computational spectrometers using colloidal quantum dots~\cite{bao_colloidal_2015}, single nanowires~\cite{yang_single-nanowire_2019}, disordered photonic chips~\cite{redding_compact_2013}, photonic crystal slabs~\cite{wang_single-shot_2019}, folded metasurface relays~\cite{faraji-dana_compact_2018}, electrically tunable van der Waals junctions~\cite{yoon_miniaturized_2022}, and angle-resolved or angle-multiplexed dielectric metasurfaces~\cite{cai_compact_2024,tittl_imaging-based_2018,leitis_angle-multiplexed_2019} have achieved impressive miniaturization in  detector area, device footprint, spectral resolution, and single-shot operation. Yet all employ a dispersion law fixed at fabrication, so the sampling pattern is set solely by the optics and detector layout. In contrast, our architecture decouples the sampling axis from the optic: by rotating two cascaded meta-optical elements relative to each other~\cite{bernet_adjustable_2008,iwami_demonstration_2020}, we select which wavelength is brought into focus, and the algorithm adaptively decides which wavelengths to probe based on the spectrum measured so far. This approach transfers content-adaptive sampling techniques previously explored in computational imaging~\cite{dai_adaptive_2014,yu_adaptive_2014,saragadam_wavelet_2019} and autonomous experimental platforms~\cite{noack_gaussian_2021,lawson_adaptive_2025,sullivan_active_2015} onto a compact meta-optical system, where the dispersion law is reconfigured on demand rather than assumed to be a fixed measurement matrix to be inverted numerically.



Our optical setup needs only a single photo-detector positioned behind a fixed collection aperture. Operating in this single-pixel mode can significantly lower both power consumption and overall cost. Our platform therefore holds broader promise for compact spectroscopic systems. A programmable one-dimensional degree of freedom supports spectrally informed scanning strategies and enables task-specific scanning modes, allowing dynamic switching from coarse spectral scans to fine-resolution scans for detecting narrow emission lines, depending on the application needs.

\paragraph{Limitations and Future Directions.} The current prototype has limitations primarily arising from the rotation stage.
We chose the rotation stage for its speed and simplicity; however it operates in an open loop configuration, requiring complete halt before next measurement, resulting in significant overhead between samples.
Future work can leverage precision rotation stages that are extremely fast to achieve real-time spectrometer.
Our meta-optical element currently works for visible to NIR wavelengths. Future work can leverage anti-reflective coatings, and broadband materials to implement higher throughput broadband spectrometers.

%
%
%
%

\section{Methods}\label{sys_design}

\subsection{Fabrication of meta-optics}

A SiN thin film of 1000\,nm thickness was deposited via PECVD (SPTS PECVD) on a 0.5 mm thick quartz wafer. After deposition, the wafer was diced to 1.5 cm squares. A positive resist (ZEP 520 A) was applied onto the sample with a maximum 4k rpm, which yielded a thickness of approximately 400 nm. The resist was baked on a hot plate at 180\,$^\circ$C for 3 min, followed by a cooling step and subsequent deposition of a conductive polymer layer (DisCharge H2O), with the same spin parameters. We then used electron beam lithography (EBL) with a 100 keV electron beam (JEOL JBX6300FS) at a dose of 275 $\mu$C cm-2. After EBL, the conductive polymer layer was removed using a brief IPA rinse, followed by developing of the resist in Amyl Acetate for 2 min at room temperature. Subsequently, we used a brief oxygen plasma descum step in a barrel etcher (100 W, 15\,s). Following, a layer of 100\,nm AlOx was deposited using electron beam evaporation. The deposited layer was lifted off overnight in a heated NMP bath. After further cleaning in Acetone/IPA and barrel etcher, the mask design was transferred into the underlying SiN film in an ICP RIE etcher using a mixture of C4F8/SF6 (Oxford PlasmaLab System 100). 

\subsection{Hardware details}\label{hardware}

The spectrometer comprises of two cascaded meta-optics, MO$_1$ and MO$_2$, mounted on individual translation and rotation stages along a common optical axis. MO$_1$ is the rotating element, seated on a piezo-motorized rotation stage, while MO$_2$ is its static counterpart, and both are mounted on lateral translation stages that allow the centroids of the two meta-optics to be brought onto the rotation axis to high precision. The wavelength-selective focusing of the cascaded pair is extremely sensitive to lateral decenter and tilt between the two elements~\cite{iwami_demonstration_2020,ogawa_rotational_nodate,li_large-gap_2025}, so we use a PSF-driven visual procedure to enforce co-axiality between MO$_1$ and MO$_2$, registration of MO$_1$ to the rotation axis, and minimal tilt between the two surfaces and the sensor plane; the full procedure is described in Supplementary Note~4 and illustrated in Supplementary Figs.~6 and 7. Illumination is provided by a fiber-coupled, approximately point-like source placed at the $2f$ distance from MO$_1$, and the focus formed by the cascaded pair is captured by a monochrome scientific camera operated in 16-bit mode with $4 \times 4$ hardware binning, which sums the charge from 16 native pixels into each binned readout pixel before digitization and matches the spatial extent of
the binned pixel to that of the focused PSF spot, raising the per-readout signal-to-noise ratio. For spectroscopic acquisition only the single binned pixel coinciding with the on-axis PSF centroid is read out at every angle, with ten frames averaged at a 5\,ms exposure to suppress temporal noise, so the per-angle output of the system is one scalar intensity value and the detector functions as a single-pixel photo-detector~\cite{edgar_principles_2019,duarte_single-pixel_2008} with a one-dimensional spectral reconstruction.

\subsection{Adaptive Faber-Schauder bisection}
\label{sec:adaptive-scanning}

The content-adaptive scanner of Section~\ref{subsec:content-adaptive-spectrometry} is implemented as a Faber-Schauder bisection algorithm on the rotation axis, following
adaptive bisection refinement in computational imaging~\cite{saragadam_wavelet_2019,dai_adaptive_2014,yu_adaptive_2014}, active wavelength selection in tunable
sensing~\cite{sullivan_active_2015}, and Gaussian-process autonomous experiments at large-scale facilities~\cite{noack_gaussian_2021,lawson_adaptive_2025}.

\paragraph{Detail coefficient.}
For each leaf interval $[x_{L}, x_{R}]$ with midpoint $x_{M}$, the local detail is scored by the Faber-Schauder coefficient
\begin{equation}
    d \;=\; \left| f(x_{M}) - \tfrac{1}{2}\bigl(f(x_{L}) + f(x_{R})\bigr) \right|,
    \label{eq:fs-detail}
\end{equation}
i.e., the local second-order finite difference of $f$ on $[x_{L}, x_{R}]$. This is the natural coefficient in the dyadic Faber-Schauder hat-function expansion of piecewise-linear
functions~\cite{douzi_faber-schauder_2001,mallat_wavelet_1999} and a proxy for the local Haar wavelet magnitude; it vanishes on locally affine regions and grows quadratically with interval width on regions of high curvature.

\paragraph{Refinement and reconstruction.}
Intervals whose detail exceeds an intensity-normalized threshold are bisected by acquiring a single new measurement at the midpoint, subject to a top-$K$ per-round splitting budget~\cite{berger_adaptive_1984} and to hardware guards on the minimum interval width $\Delta\theta_{\min}$ and the maximum tree depth $D_{\max}$. Within each round, midpoints are visited in nearest-neighbor order from the current stage position to minimize total angular travel without affecting reconstruction fidelity. The spectrum is reconstructed by piecewise-linear interpolation on the resulting non-uniform mesh; the estimator is {\em anytime}, improves monotonically with samples, and is consistent with the optimal $N$-term rate for piecewise-smooth signals on adaptive dyadic partitions~\cite{devore_nonlinear_1998,cohen_adaptive_2001,baraniuk_model-based_2010}.

The full algorithmic partition which includes pilot length, threshold normalization, per-round budget, termination conditions are discussed in Supplementary Note~2 (Supplementary Algorithm~1), with all parameter values in Supplementary Table~1.

\section{Acknowledgements}
This work was supported by NSF grants NSF-2120774, CCF-2403123, and DTRA grant HDTRA12610007. Part of this work was conducted at the Washington Nanofabrication Facility / Molecular Analysis Facility, a National Nanotechnology Coordinated Infrastructure (NNCI) site at the University of Washington with partial support from the National Science Foundation via awards NNCI‑1542101 and NNCI‑2025489.

\bibliography{sn-bibliography}

\end{document}


\title[Article Title]{Supplementary Material}




\affil*[1]{\orgdiv{Department of Electrical Engineering}, \orgname{University of California,Riverside}, \orgaddress{\street{900 University Ave}, \city{Riverside}, \postcode{92507}, \state{California}, \country{USA}}}

\affil[2]{\orgdiv{Department of Electrical and Computer Engineering}, \orgname{University of Washington}, \orgaddress{\street{185 Stevens Way}, \city{Seattle}, \postcode{10587}, \state{Washington}, \country{USA}}}





\maketitle




\section{Derivations}
\label{supp:derivations}

This section develops the fundamental theory of the Moir\'e meta-spectrometer. We begin by deriving the dispersion law used throughout the main text, demonstrating an affine relationship between the wavenumber and the relative rotation angle of the two Moir\'e metasurfaces. We then show that, under uniform angular sampling, the corresponding spectral interval $\Delta\lambda$ scales as $\lambda^{2}$. This scaling arises directly from the linear dispersion in wavenumber and serves as the geometric basis for the content-adaptive scanner.

\subsection{Moir\'e Metalens Principle and Angle-Wavelength Calibration}

The operational principle of Moir\'e meta-optics relies on the mutual rotation of two complementary metasurfaces~\cite{bernet_adjustable_2008,bernet_demonstration_2013}. We describe the phase profiles of the two metasurfaces in polar coordinates $(r,\phi)$ as $\Phi_{\MO_{1}}(r,\phi)$ and $ \Phi_{\MO_{2}}(r,\phi)$, with corresponding complex transmission functions
\begin{equation}
T_{\MO_{1}}(r,\phi)=\exp\!\left(i\Phi_{\MO_{1}}(r,\phi)\right),\qquad
T_{\MO_{2}}(r,\phi)=\exp\!\left(i \Phi_{\MO_{2}}(r,\phi)\right).
\end{equation}
To enable Moir\'e tuning, the metasurfaces are designed to be phase-complementary, such that $ \Phi_{\MO_{2}}(r,\phi)=-\Phi_{\MO_{1}}(r,\phi)$. Rotating the second metasurface by a relative angle $\theta$ modifies its transmission to $T_{\MO_{2}}(r,\phi-\theta)=\exp\!\left(-i\Phi_{\MO_{1}}(r,\phi-\theta)\right)$. The joint transmission of the cascaded stack is therefore
\begin{equation}
T_{\mathrm{joint}}(r,\phi;\theta)=T_{\MO_{1}}(r,\phi)\,T_{\MO_{2}}(r,\phi-\theta)
=\exp\!\left(i\left[\Phi_{\MO_{1}}(r,\phi)-\Phi_{\MO_{1}}(r,\phi-\theta)\right]\right).
\label{eq:joint_phase}
\end{equation}
For small relative rotations, the phase difference in Eq.~\eqref{eq:joint_phase} admits a first-order Taylor approximation,
\begin{equation}
\Phi_{\MO_{1}}(r,\phi)-\Phi_{\MO_{1}}(r,\phi-\theta)\approx \theta\,\frac{\partial \Phi_{\MO_{1}}(r,\phi)}{\partial \phi},
\label{eq:taylor}
\end{equation}
which shows that rotation converts an angular phase gradient into an effective, rotation-controlled wavefront.

To synthesize a tunable lens, we adopt a separable phase design $\Phi_{\MO_{1}}(r,\phi)=\Phi_r(r)\,\Phi_\phi(\phi)$ and seek a net quadratic phase that matches a thin lens of focal length $f$ at wavelength $\lambda$,
\begin{equation}
T_{\mathrm{lens}}(r)=\exp\!\left(-i\frac{\pi r^2}{\lambda f}\right).
\label{eq:standard_lens}
\end{equation}
Choosing $\Phi_r(r)=a r^2$ and $\Phi_\phi(\phi)=\phi$ yields $\partial \Phi_{\MO_{1}}(r,\phi)/\partial \phi = a r^2$. Substituting into Eq.~\eqref{eq:taylor} and equating to the lens phase in Eq.~\eqref{eq:standard_lens} (the overall sign is fixed by the rotation sense, equivalently by the sign of $a$) gives the matching condition
\begin{equation}
a r^2\,\theta=\frac{\pi r^2}{\lambda f}.
\label{eq:matching_condition}
\end{equation}

%

Accounting for a constant rotation stage offset compared to the Moir\'e lens pair,
\begin{equation}
a r^2\,(\theta-\theta_0)=\frac{\pi r^2}{\lambda f},
\label{eq:matching_condition_offset}
\end{equation}
which directly yields the calibration relationship
\begin{equation}
\theta=\left(\frac{\pi}{a f}\right)\frac{1}{\lambda}+\theta_0.
\label{eq:calibration}
\end{equation}
Equation~\eqref{eq:calibration} predicts a monotonic inverse dependence between rotation angle and wavelength for fixed $f$ (i.e., $\theta\propto \lambda^{-1}$), with $\theta_0$ capturing the offset between the optical and mechanical angular references. Experimentally, we calibrate this mapping using a set of narrow-linewidth laser lines (Supplementary Note~\ref{sec:methods_all}) to associate each wavelength with a unique rotation angle read by the encoder, enabling wavelength assignment and spectral reconstruction from a controlled rotational scan.

\subsection{Wavelength-squared scaling of the spectral interval}
\label{ssec:resolution}

The calibrated dispersion law of Eq.~\eqref{eq:calibration} is affine in wavenumber $\nu \equiv 1/\lambda$, with slope $d\theta/d\nu = \pi/(a f)$ independent of $\lambda$. Equal commanded angular steps $\Delta\theta$ therefore correspond to equal wavenumber steps,
\begin{equation}
  \Delta\nu \;=\; \frac{a f}{\pi}\,\Delta\theta,
\end{equation}
and a uniform sweep in $\theta$ is a uniform sweep in $\nu$.
Converting to a wavelength interval via $\nu = 1/\lambda$ gives
$\Delta\nu = -\Delta\lambda/\lambda^{2}$, so
\begin{equation}
  \Delta\lambda(\lambda) \;=\; \frac{a f \lambda^{2}}{\pi}\,\Delta\theta.
  \label{eq:dlambda}
\end{equation}
The $\lambda^{2}$ scaling is therefore a direct geometric consequence of dispersing linearly in wavenumber while sampling uniformly in angle, and is independent of any property of the focusing optic beyond the dispersion law itself. A fixed angular step delivers finer $\Delta\lambda$ at short wavelengths and coarser $\Delta\lambda$ at long wavelengths, naturally matching the wavelength dependence of typical atomic and molecular line densities.

\section{Adaptive Scanning Algorithm}
\label{supp:adaptive}
 
This section details the content-adaptive Faber–Schauder scanner used throughout the main paper. The acquisition proceeds in three stages: a content-agnostic coarse pilot sweep that initializes the interval tree (Phase 1, Section~\ref{supp:adaptive:pilot}); a budget-limited, iterative refinement loop that repeatedly bisects intervals containing high levels of detail (Phase 2, Section~\ref{supp:adaptive:refinement}); and a final piecewise-linear reconstruction defined on the resulting non-uniform mesh (Phase 3, Section~\ref{supp:adaptive:reconstruction}). The full procedure is summarized in Algorithm~S\ref{alg:wtp-full}, and the parameter choices used in this work are listed in
Table~S\ref{tab:wtp-params}.
 
\subsection{Notation and the Faber--Schauder detail coefficient}
\label{supp:adaptive:detail}
Let $f(\theta)$ denote the measured intensity at rotation angle $\theta$, and $\hat{f}(\theta)$ its piecewise-linear reconstruction from a finite sample set. The acquisition spans the angular domain $[\theta_{\min}, \theta_{\max}]$, which
Eq.~\eqref{eq:calibration} maps bijectively to the wavelength axis $\lambda(\theta)$. For a subinterval $[\theta_L, \theta_R]$ with midpoint $\theta_M = (\theta_L + \theta_R)/2$, the linear interpolant predicts a midpoint intensity $\tilde{f}(\theta_M) = \tfrac12(f_L + f_R)$. After measuring $f_M = f(\theta_M)$, the deviation from this prediction defines the Faber--Schauder detail coefficient
\begin{equation}
    d[\theta_L, \theta_R] \;\triangleq\;
    \bigl|\, f_M - \tfrac12(f_L + f_R)\,\bigr|,
    \label{eq:supp-fs-detail}
\end{equation}
the local second-order finite difference of $f$. Three properties make $d$ a well-founded refine-or-stop criterion. First, it vanishes on affine intervals: if
$f$ is linear over $[\theta_L, \theta_R]$ then $d = 0$, so smooth, slowly varying
regions yield negligible detail and stop refining early. Second, it scales
quadratically with the interval width and tracks the local curvature of $f$; for
$f \in C^2$, a Taylor expansion about $\theta_M$ yields
\begin{equation}
    d[\theta_L, \theta_R] = \tfrac{h^2}{4}\,\bigl|f''(\xi_L) + f''(\xi_R)\bigr|
    \;\le\; \tfrac{h^2}{2}\,\|f''\|_{\infty,[\theta_L,\theta_R]},
    \label{eq:supp-fs-taylor}
\end{equation}
where $h = (\theta_R - \theta_L)/2$ is the half-interval width and $\xi_L, \xi_R$
lie in the left and right sub-intervals. The detail is therefore largest near sharp
emission lines and decays as $h^2$ once the interval is small enough for the
interpolant to be accurate, concentrating samples where the spectrum is sharp.
Third, $d$ is the magnitude of a wavelet coefficient rather than an empirical
heuristic: expanding $f$ in the dyadic Faber--Schauder hat-function basis,
\begin{equation}
    f(\theta) = f(\theta_{\min})\,\varphi_{0,0}(\theta)
    + f(\theta_{\max})\,\varphi_{0,1}(\theta)
    + \sum_{\ell \ge 0}\sum_{k} c_{\ell,k}\,\psi_{\ell,k}(\theta),
    \label{eq:supp-fs-expansion}
\end{equation}
where $c_{\ell,k} = f(\theta_{\ell,k,M}) - \tfrac12\bigl[f(\theta_{\ell,k}) +
f(\theta_{\ell,k+1})\bigr]$ is the signed form of Eq.~\eqref{eq:supp-fs-detail}.
The scanner's decision metric $d$ is thus exactly the absolute Faber--Schauder
coefficient of $f$ on the corresponding dyadic
interval~\cite{douzi_faber-schauder_2001,mallat_wavelet_1999}.

\subsection{ Phase 1: Pilot sweep}
\label{supp:adaptive:pilot}
 
Acquisition begins with $N_{\mathrm{pilot}}$ equally spaced angular
positions across $[\theta_{\min}, \theta_{\max}]$. We use
$N_{\mathrm{pilot}} = 5$ throughout this work; this choice is small
enough to leave most of the sample budget for the adaptive phase yet
large enough that every level-zero interval has a chance of detecting
a peak of width $\gtrsim 1\%$ of the sweep range. The five pilot
positions partition the domain into $N_{\mathrm{pilot}} - 1 = 4$
level-zero intervals, which form the initial refinement frontier
$\mathcal{F}^{(0)}$.
 
To minimise unnecessary motor travel during the pilot, the five
positions are visited in greedy nearest-neighbour order starting from
the current stage position $\theta_{\mathrm{curr}}$:
\begin{equation}
    \theta_{\mathrm{next}}
    \;=\;
    \mathop{\arg\min}_{\theta \in S_{k}}\,
    \bigl|\,\theta - \theta_{\mathrm{prev}}\,\bigr|,
    \label{eq:supp-greedy}
\end{equation}
where $S_{k}$ is the set of yet-unvisited pilot positions at step
$k$. On a one-dimensional domain this heuristic is optimal whenever
$\theta_{\mathrm{curr}}$ lies outside the convex hull of $S_{0}$ and
incurs at most a $2\times$ overhead otherwise.

\subsection{Phase 2: Iterative refinement}
\label{supp:adaptive:refinement}
 
Refinement proceeds in successive rounds $t = 1, \dots, R$. Each
round comprises four steps.
 
\paragraph{Step 1 (filter).}
The current frontier $\mathcal{F}^{(t-1)}$ is filtered to a candidate
set $\mathcal{C}^{(t)}$ by removing nodes that have either reached
the maximum tree depth $D_{\max}$ or whose interval width has
dropped below the minimum reproducible motor step
$\Delta\theta_{\min}$. Nodes failing either guard are retained on the
frontier but are not eligible for splitting in subsequent rounds.
 
\paragraph{Step 2 (pre-measurement of midpoints).}
For each candidate interval $[\theta_{L}, \theta_{R}] \in \mathcal{C}^{(t)}$ whose midpoint has not yet been probed, the spectrometer records a single measurement at $\theta_{M} = (\theta_{L} + \theta_{R})/2$. If this midpoint (within a small angular tolerance) coincides with an already sampled position, the value is retrieved from cache instead of being remeasured. The detail coefficient $d[\theta_{L}, \theta_{R}]$ is then evaluated using Eq.~(\ref{eq:supp-fs-detail}). Every candidate must be measured before ranking: ranking intervals only by the endpoint difference $|f_{R} - f_{L}|$ would often miss symmetric structures such as emission lines, whose endpoints can be nearly identical despite a strong peak in between.
 
\paragraph{Step 3 (intensity-normalised threshold).}
The detail threshold for round $t$ is set relative to the running
peak intensity,
\begin{equation}
    \tau_{\mathrm{rel}}^{(t)}
    \;=\;
    \tau \cdot \max_{\theta' \in \mathcal{S}^{(t)}}\,f(\theta'),
    \label{eq:supp-threshold}
\end{equation}
where $\mathcal{S}^{(t)}$ is the set of all angular positions
measured up to and including round $t$, and $\tau \in (0, 1)$ is the
single dimensionless control parameter of the scanner. Because the
threshold is rescaled by the running peak rather than fixed in
absolute camera units, $\tau$ behaves as a fractional detail-floor:
$\tau = 10^{-3}$, for example, means ``split any interval whose
midpoint prediction error exceeds $0.1\%$ of the largest measured
intensity so far.'' The self-normalising form has the additional
property that it absorbs source-brightness variation across
experiments without requiring per-source recalibration.
 
\paragraph{Step 4 (top-$K$ split, with re-queue).}
The candidate set $\mathcal{C}^{(t)}$ is sorted by detail
coefficient in descending order. A budgeted top-$K$ policy then
walks the sorted list and:
\begin{itemize}
\item[(a)] declares an interval a \emph{leaf} (and removes it from
the frontier) if its detail is below $\tau_{\mathrm{rel}}^{(t)}$ or
if its bisected child width would fall below $\Delta\theta_{\min}$;
\item[(b)] \emph{bisects} the interval if its detail exceeds
$\tau_{\mathrm{rel}}^{(t)}$, by acquiring a new sample at the
midpoint (or reusing the pre-measured midpoint from Step~2) and
creating two child intervals $[\theta_{L}, \theta_{M}]$ and
$[\theta_{M}, \theta_{R}]$ at depth $\ell + 1$;
\item[(c)] \emph{re-queues} the interval onto the next round's
frontier $\mathcal{F}^{(t)}$ once the per-round split count reaches
$K$, so that high-detail intervals that did not fit in the current
round's budget are reconsidered in the next round.
\end{itemize}
The top-$K$ budget is structurally analogous to top-down adaptive
mesh refinement~\cite{berger_adaptive_1984}: it prevents any single
round from spending the entire acquisition budget on one feature
and ensures that detail is distributed across multiple peaks before
any one peak is refined to its depth cap.
 
\paragraph{Termination.}
The refinement loop exits when any of four conditions holds:
the frontier is empty (every interval has been declared a leaf);
the candidate set is empty (all remaining intervals have hit
$D_{\max}$ or $\Delta\theta_{\min}$); the round counter reaches the
configured cap $R$; or the total sample count reaches the budget $B$.
On termination, the leaf intervals
$\mathcal{L}^{(T)} = \{[\theta_{i}, \theta_{i+1}]\}$ form a
non-uniform partition of $[\theta_{\min}, \theta_{\max}]$.
 
The full algorithm is given in Algorithm~S\ref{alg:wtp-full}.
 
\begin{algorithm}[t]
\caption{Content-Adaptive Faber--Schauder Scanner.}
\label{alg:wtp-full}
\begin{algorithmic}[1]
\Require domain $[\theta_{\min},\theta_{\max}]$, relative threshold $\tau$, sample budget $B$;
  guards: max depth $D_{\max}$, min width $\Delta\theta_{\min}$, splits/round $K$, max rounds $R$.
\Ensure samples $\mathcal{S}=\{(\lambda_i,f_i)\}$ and piecewise-linear reconstruction $\hat f(\lambda)$.
\Statex
\State \textbf{Pilot:} measure $N_{\mathrm{pilot}}$ equally spaced angles in nearest-neighbour
  order (Eq.~\ref{eq:supp-greedy}), store in $\mathcal{S}$, and seed the frontier $\mathcal{F}$
  with the consecutive intervals at depth $0$.
\Statex
\For{$t=1,\dots,R$}\Comment{refinement rounds}
  \If{$|\mathcal{S}|\ge B$ \textbf{or} $\mathcal{F}=\varnothing$} \State \textbf{break} \EndIf
  \State $\mathcal{C}\gets\{\,I=[\theta_L,\theta_R]\in\mathcal{F}:
    \operatorname{depth}(I)<D_{\max},\ \theta_R-\theta_L\ge\Delta\theta_{\min}\,\}$
  \ForAll{$I\in\mathcal{C}$}\Comment{probe midpoints}
    \State $m\gets\tfrac12(\theta_L+\theta_R)$;\ measure $f(m)$ if not cached
    \State $d_I\gets\bigl|f(m)-\tfrac12(f(\theta_L)+f(\theta_R))\bigr|$\Comment{Faber--Schauder detail}
  \EndFor
  \State $\tau_{\mathrm{rel}}\gets\tau\cdot\max_{\theta\in\mathcal{S}}f(\theta)$;\quad
    $\mathcal{F}\gets\varnothing$;\quad $k\gets0$
  \ForAll{$I\in\mathcal{C}$ in order of decreasing $d_I$}
    \If{$d_I<\tau_{\mathrm{rel}}$ \textbf{or} child width $<\Delta\theta_{\min}$}
      \State \textbf{continue}\Comment{interval resolved}
    \ElsIf{$k<K$}
      \State add children $[\theta_L,m],[m,\theta_R]$ to $\mathcal{F}$ at depth $+1$;\ $k\gets k+1$
    \Else
      \State re-queue $I$ to $\mathcal{F}$\Comment{carried to next round}
    \EndIf
  \EndFor
\EndFor
\Statex
\State \textbf{Reconstruct:} map $\theta_i\mapsto\lambda_i$ (Eq.~2) and interpolate $\hat f(\lambda)$
  piecewise-linearly over $\mathcal{S}$ sorted by $\lambda$.
\State \Return $\mathcal{S},\ \hat f(\lambda)$
\end{algorithmic}
\end{algorithm}

\subsection{Phase 3: Reconstruction}
\label{supp:adaptive:reconstruction}
 
After termination, the cached samples
$\mathcal{S} = \{(\theta_{i}, f(\theta_{i}))\}$ are sorted in
ascending order of angle and converted to wavelength. Within each leaf interval
$[\theta_{i}, \theta_{i+1}]$ the spectrum is reconstructed by
piecewise-linear interpolation,
\begin{equation}
    \hat{f}(\theta)
    \;=\;
    f(\theta_{i})
    + \frac{f(\theta_{i+1}) - f(\theta_{i})}{\theta_{i+1} - \theta_{i}}\,
      (\theta - \theta_{i}),
    \qquad \theta \in [\theta_{i}, \theta_{i+1}],
    \label{eq:supp-recon}
\end{equation}
and the result is mapped to the wavelength axis to give
$\hat{f}(\lambda)$. The estimator is \emph{anytime}: a usable
reconstruction is available after every round of refinement, and the
sequence $\hat{f}_{1}, \hat{f}_{2}, \dots$ converges monotonically as
$|\mathcal{S}|$ grows. For functions of bounded second variation,
piecewise-linear interpolation on an adaptively refined dyadic mesh
achieves the optimal $N$-term approximation rate among order-one
schemes~\cite{devore_nonlinear_1998,cohen_adaptive_2001}, which
matches the regularity class of typical spectra: locally smooth
across broad continua, locally rough at narrow emission lines.

\subsection{Parameter values used in this work}
\label{supp:adaptive:params}
 
Table~S\ref{tab:wtp-params} lists the control parameters used for
every measurement in the main text and supplementary figures.
The same parameter set was applied to every light source; no
per-source retuning was performed.
 
\begin{table}[h]
  \centering
  \caption{Parameters of the content-adaptive Faber--Schauder scanner.
    These are the fixed configuration used for the adaptive-versus-uniform
    benchmark of Supplementary
    Figs.~\ref{fig:psnr_vs_sampling_budget}--\ref{fig:psnr_vs_acq_time};
    the sample budget $B$ is swept over the listed values, and the detail
    threshold $\tau$ is the only fidelity parameter (the remaining entries
    are hardware-aware guards). The minimum interval width
    $\Delta\theta_{\min}$ is a software floor on bisection, not the physical
    stage resolution.}
  \label{tab:wtp-params}
  \begin{tabular}{l l l l}
    \toprule
    Symbol & Parameter & Setting key & Value \\
    \midrule
    $[\theta_{\min}, \theta_{\max}]$ & Angular sweep range
      & \texttt{SWEEP\_PARAMETERS} & $[-10^{\circ},\,+40^{\circ}]$ \\
    $N_{\mathrm{pilot}}$ & Pilot sample count
      & \texttt{N\_COARSE} & $5$ \\
    $\tau$ & Detail threshold (fractional)
      & \texttt{wtp\_tau} & $10^{-4}$ \\
    $D_{\max}$ & Maximum tree depth
      & \texttt{wtp\_max\_depth} & $30$ \\
    $\Delta\theta_{\min}$ & Minimum interval width (software guard)
      & \texttt{wtp\_min\_width\_deg} & $0.2^{\circ}$ \\
    $K$ & Splits per round (top-$k$)
      & \texttt{wtp\_top\_k} & $1$ \\
    $R$ & Maximum refinement rounds
      & \texttt{wtp\_max\_oscillations} & $100$ \\
    $B$ & Total sample budget (swept)
      & \texttt{SAMPLING\_BUDGET} & $\{8,16,32,64,128,256\}$ \\
    $t_{\mathrm{settle}}$ & Settle time per measurement
      & \texttt{settle\_seconds} & $10$\,ms \\
    \bottomrule
  \end{tabular}
\end{table}
 
The single dimensionless control parameter $\tau$ mediates the trade-off between sample efficiency and reconstruction fidelity. Smaller values of $\tau$ allow more sub-threshold features to be captured and drive more aggressive refinement, yielding higher-fidelity reconstructions at the expense of additional measurements. Larger values, in contrast, enforce stronger pruning and earlier termination, resulting in coarser reconstructions that require fewer samples. Empirically, we observe that $\tau \in [10^{-6},\,10^{-4}]$ spans the range in which the adaptive scanner surpasses a uniform sweep at matched fidelity for all tested source classes. The remaining parameters ($D_{\max}, \Delta\theta_{\min}, K, R$) are hardware-constrained limits rather than fidelity controls: $\Delta\theta_{\min}$ is determined by the minimum reproducible increment of the ELL14 rotation stage, $D_{\max}$ and $R$ act as safeguards against pathological inputs, and $K = 2$ guarantees that fine structure is shared across multiple peaks before any individual peak is refined to its maximum depth.

 

\section{Adaptive Scan Results}
\label{supp:adaptive_scan_results}

\begin{figure*}[t]
  \centering
  \includegraphics[width=1.0\linewidth]{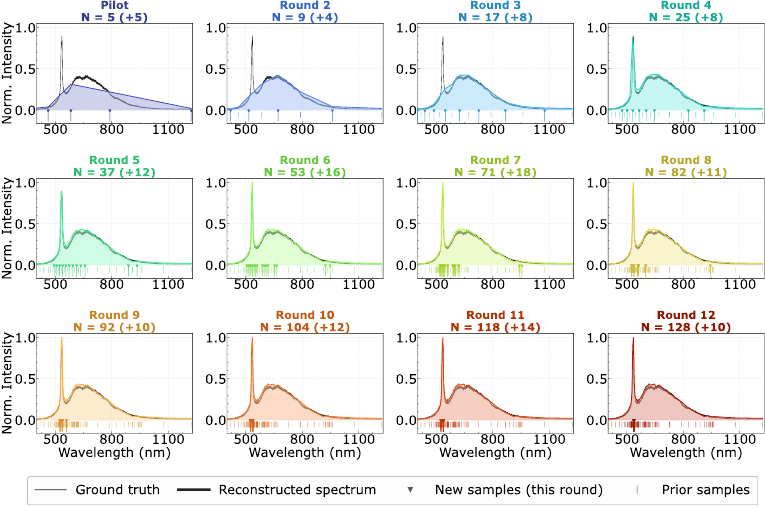}
  \caption{\textbf{Round-by-round adaptive reconstruction of a broadband source with an embedded laser line.} Each panel shows the piecewise-linear reconstruction $\hat f_t(\lambda)$ (colored fill) on the cumulative sample set at the end of round $t$, with the 0.01$^\circ$-step ground-truth scan overlaid in black. Panels are labeled by round (Pilot, Round 2--12) and annotate the cumulative sample count $N$ together with the new samples added in that round (in parentheses). Tick marks beneath each axis indicate the angular positions of the new samples (colored) and prior samples (grey). The pilot sweep places $N=5$ uniform samples across the full band; subsequent rounds preferentially refine the high-curvature region around the laser line while leaving the smooth continuum under-sampled, recovering both features by round~12 with $N=128$ total measurements.}
  \label{fig:exp-scan-all-rounds-mixed}
\end{figure*}

\begin{figure*}[t]
  \centering
  \includegraphics[width=1.0\linewidth]{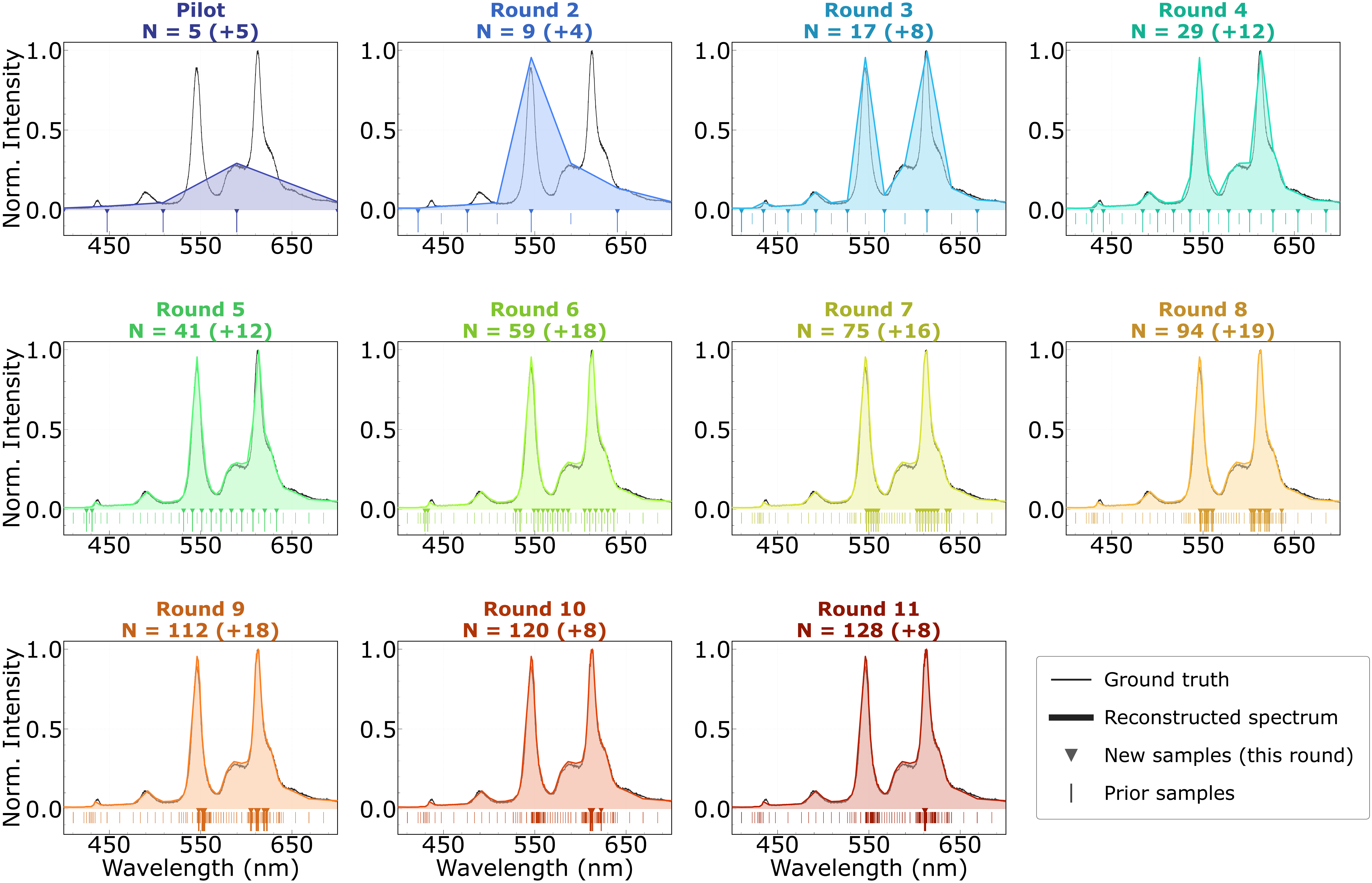}
  \caption{\textbf{Round-by-round adaptive reconstruction of a compact fluorescent lamp (CFL).} Same convention as Supplementary~Fig.~\ref{fig:exp-scan-all-rounds-mixed}. The CFL emission combines sharp mercury lines near 436, 546, 578, and 611~nm with a broad phosphor continuum spanning the visible band. The pilot sweep ($N=5$) provides only a coarse continuum estimate; from Round~2 onward, the adaptive sampler concentrates new measurements at the wavelengths of the mercury lines, locking in their positions within the first few rounds and progressively narrowing their reconstructed widths toward the ground-truth FWHM. By the final round, every mercury line is resolved and the phosphor continuum is faithfully recovered with a cumulative budget of $N=128$ samples.}
  \label{fig:exp-scan-all-rounds-cfl}
\end{figure*}

\begin{figure*}[t]
  \centering
  \includegraphics[width=1.0\linewidth]{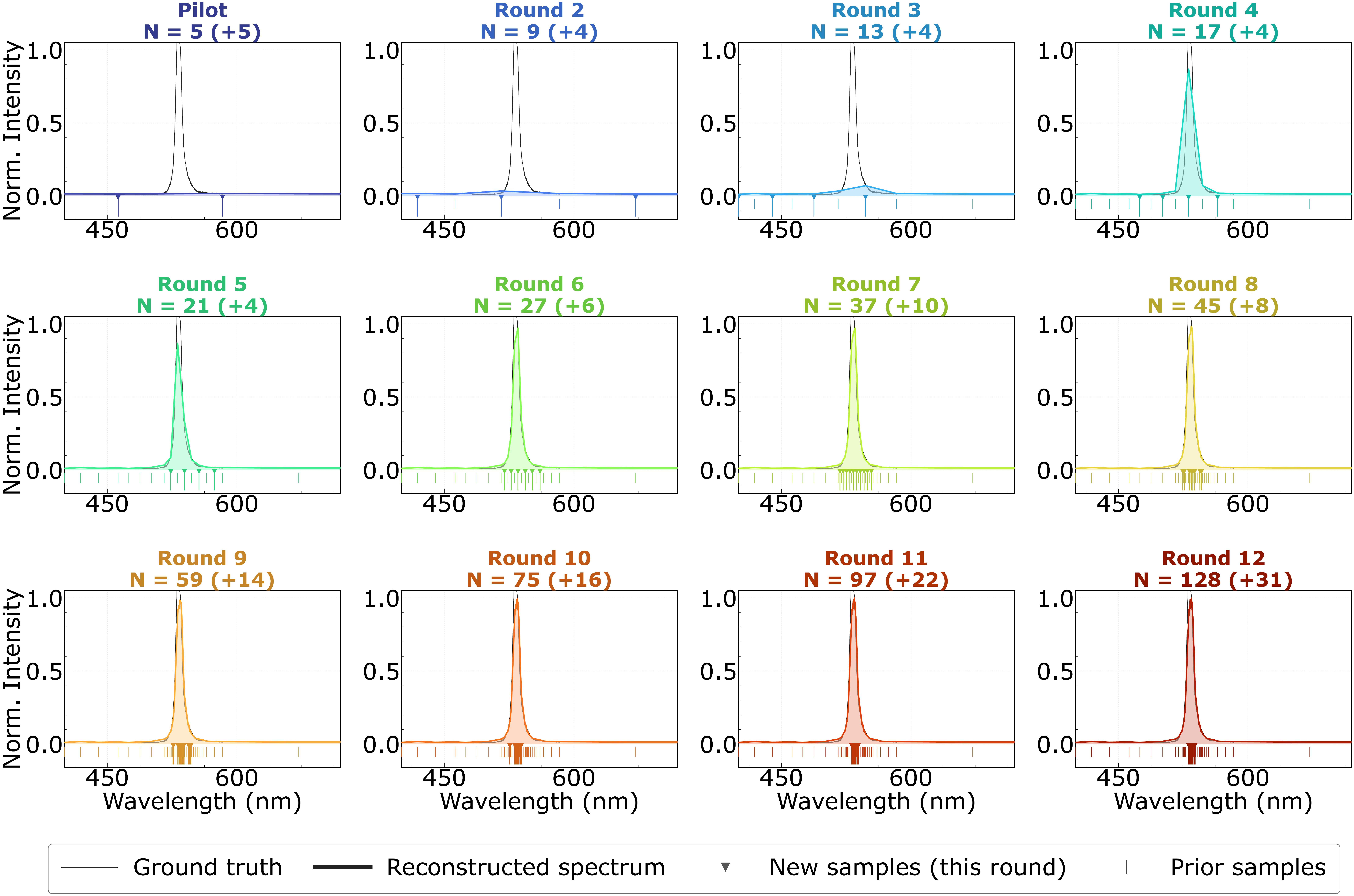}
  \caption{\textbf{Round-by-round adaptive reconstruction of a 532~nm green laser.} Same convention as Supplementary~Fig.~\ref{fig:exp-scan-all-rounds-mixed}. The source is a single
  near-monochromatic line at 532~nm --- the hardest case for blind sampling, as the $N=5$ pilot straddles it and recovers almost none of its amplitude. From Round~2 onward the adaptive sampler localises the line and spends essentially all new measurements there, restoring the peak within a few rounds and narrowing the reconstructed width toward the ground-truth FWHM until it is fully resolved at the cumulative budget $N=128$.}
  \label{fig:exp-scan-all-rounds-green}
\end{figure*}

 Supplementary Figures~\ref{fig:exp-scan-all-rounds-mixed}--\ref{fig:exp-scan-all-rounds-green} replay the content-adaptive acquisition on three sources that span the spectral-difficulty range of the main paper: a broadband continuum carrying a single embedded laser line (Fig.~\ref{fig:exp-scan-all-rounds-mixed}), a compact fluorescent lamp combining several sharp mercury lines with a phosphor continuum (Fig.~\ref{fig:exp-scan-all-rounds-cfl}), and a near-monochromatic 532~nm laser (Fig.~\ref{fig:exp-scan-all-rounds-green}). In every case the five-sample pilot resolves only the smooth envelope, and each subsequent round spends its budget where the Faber--Schauder detail is largest (on the line cores and continuum shoulders) while leaving the flat regions sparsely sampled. By the final round all sharp features are recovered at a cumulative budget of $N=128$ samples, the fidelity a uniform sweep reaches only with substantially more measurements. These experiments are the source-resolved manifestation of the main result: adaptive scanning reconstructs mixed line-plus-continuum spectra with roughly $3\times$ fewer measurements on average ( up to $7\times$ than uniform sampling at matched fidelity, with a usable spectrum available after every round. 

\paragraph{Adaptive vs Uniform Scanning}
We compare the adaptive scanner to uniform angular sampling across eleven different sources, covering narrow-band emitters (lasers, LEDs, monochromator setpoints) as well as broadband emitters (a CFL and a combined halogen\,$+\,532$\,nm source). As ground truth $G(\lambda)$, we use a densely sampled uniform sweep over $\theta \in [-10^{\circ}, +40^{\circ}]$ ($\lambda \in [384, 1220]$\,nm), padding angles outside each source’s emission band to its dark level. All evaluations are performed in simulation: the acquisition software is run unmodified on $G$, with
each measurement returning the ground truth at a slightly perturbed angle (jitter $\sigma_\theta = 0.05^{\circ}$) and corrupted by a calibrated shot-plus-read camera noise model. For uniform sampling, we distribute a total of $N \in \{8,16,32,64,128,256\}$ measurements evenly over the angular range; for the adaptive approach, we allocate the same budgets using the scanner described in Supplementary Note~\ref{supp:adaptive} (Algorithm~S\ref{alg:wtp-full}). Both methods reconstruct the signal on the dense ground-truth grid via piecewise-linear interpolation, without any super-resolution or parametric model fitting. A single set of hyperparameters ($\tau$, top-$k$, minimum width) is used for all sources, selected by a grid search that optimizes the average performance across sources.


\begin{figure*}[!tt]
  \centering
  \includegraphics[width=1.0\linewidth]{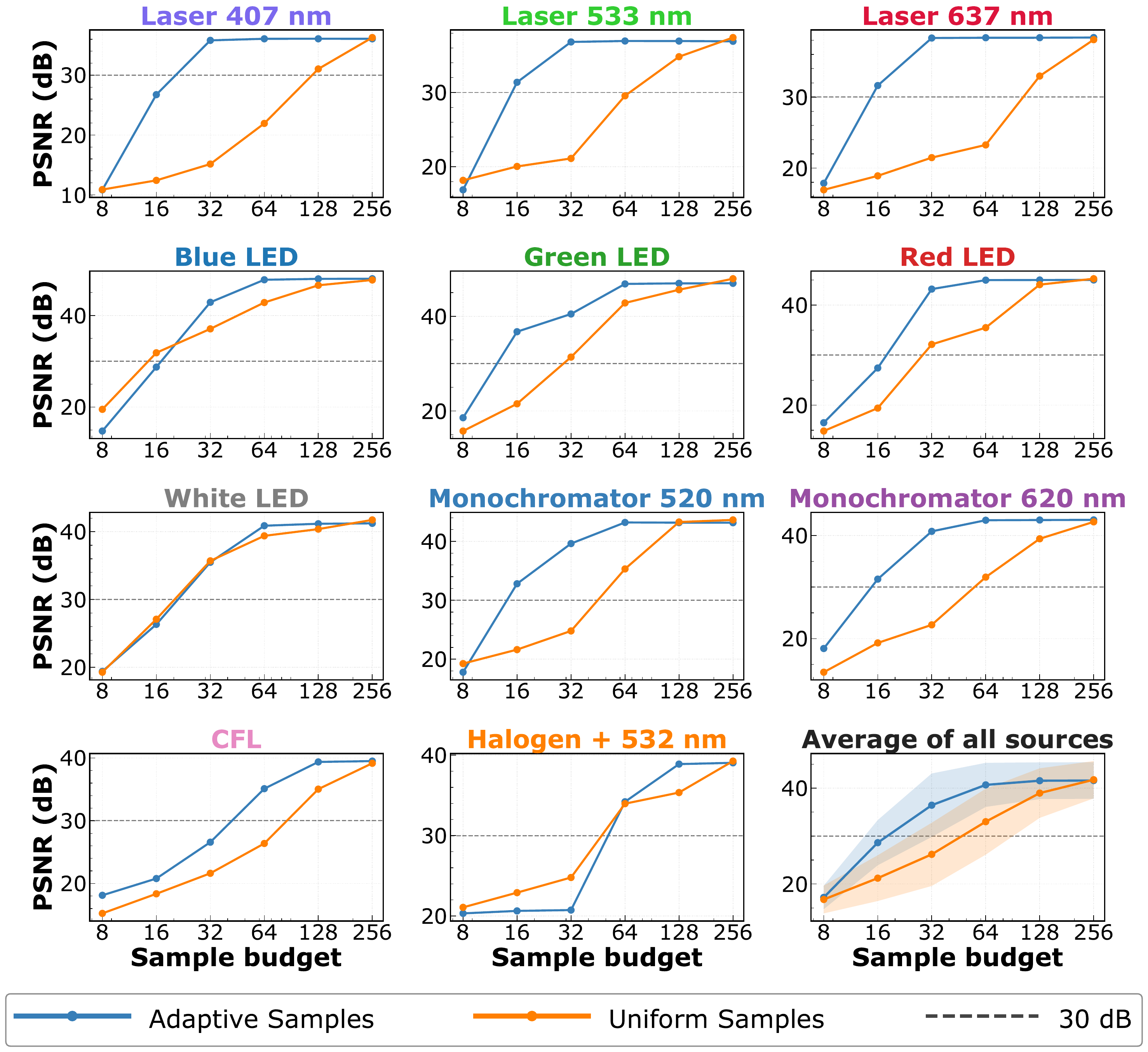}
  \caption{ \textbf{Reconstruction PSNR versus sample budget for adaptive and uniform angular sampling across eleven light sources.}
Panels show three lasers ($407$, $533$, $637$~nm), four LEDs (blue, green, red, white), two monochromator setpoints ($520$, $620$~nm), a CFL, a mixed halogen\,$+\,532$~nm source, and the across-source average (bottom right). Blue: adaptive sampling; orange: uniform sampling; dashed: $30$~dB reference. Adaptive sampling uses one fixed configuration for all sources and budgets ($5$-sample pilot, greedy top-$k=1$, detail threshold $\tau=10^{-4}$, minimum interval width $0.2^\circ$). Averaged over all sources, adaptive sampling beats uniform at every budget—by $+5.6$~dB on average and up to ${\approx}\,+11$~dB at mid budgets—with convergence as $N\!\to\!256$, where uniform becomes dense.}
  \label{fig:psnr_vs_sampling_budget}
\end{figure*}

\begin{figure*}[!tt]
  \centering
  \includegraphics[width=1.0\linewidth]{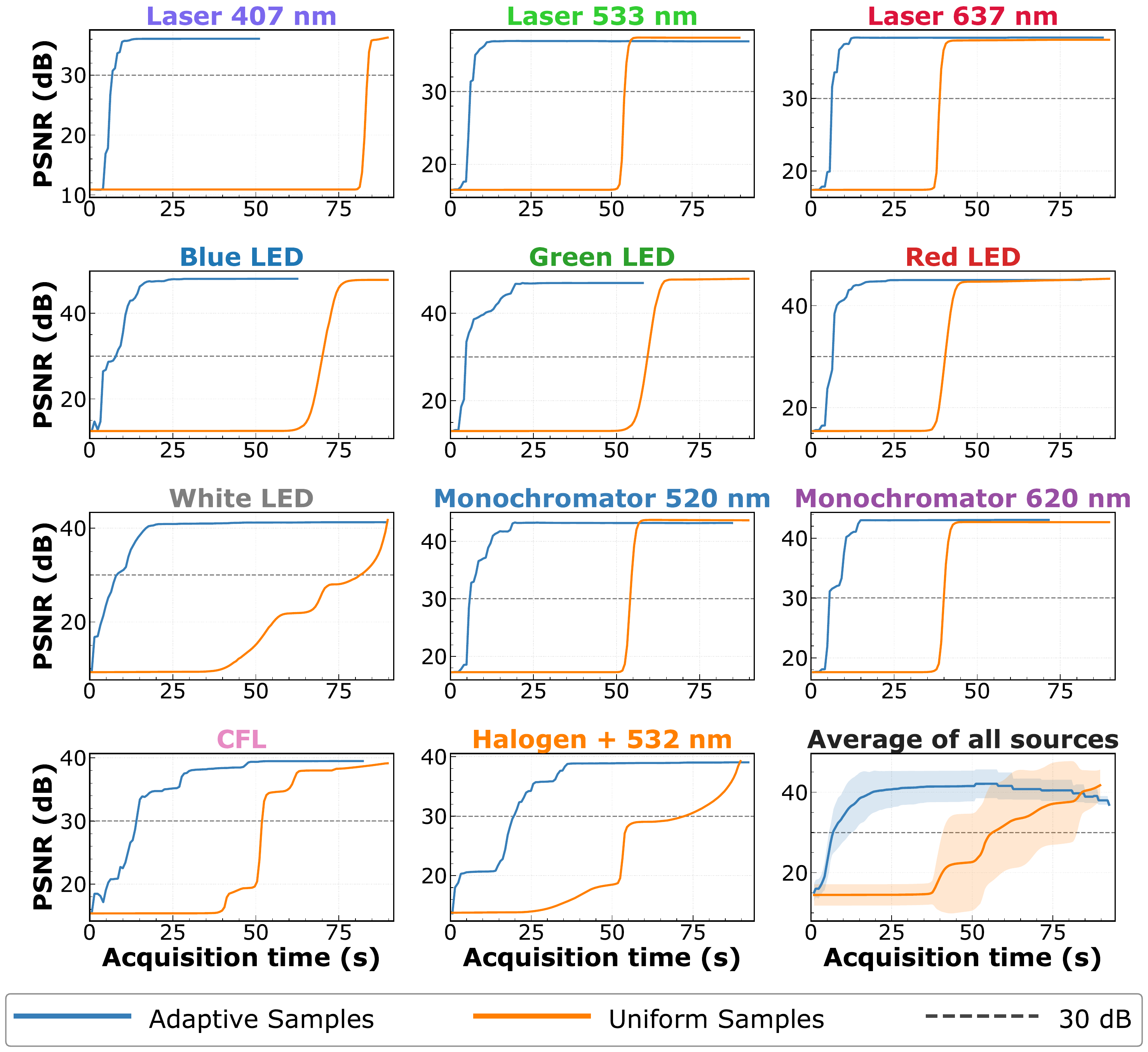}
  \caption{\textbf{Reconstruction PSNR versus acquisition time for adaptive (WTP) and
    uniform angular sampling.}
    Same eleven sources and panel layout as Fig.~\ref{fig:psnr_vs_sampling_budget}, here
    at a fixed budget of $N=256$ samples: the PSNR of the cumulative
    reconstruction is plotted against the modelled acquisition time as samples
    accrue in measurement order. Blue: adaptive sampling; orange: uniform;
    dashed line: $30$~dB. Because adaptive sampling concentrates
    measurements on spectral features instead of sweeping uniformly, it reaches
    a given PSNR in substantially less acquisition time (${\approx}\,6.7\times$
    faster to $30$~dB on average); the curves converge once uniform sampling has
    covered the full range.}
  \label{fig:psnr_vs_acq_time}
\end{figure*}

\section{Methods}\label{sec:methods_all}
This section presents the experimental realization of the Moir\'e meta-spectrometer. We outline the optomechanical construction of the cascaded meta-optic pair and its associated translation and rotation stages, the PSF-based alignment protocol used to co-register MO$_{1}$ and MO$_{2}$ along a shared optical axis, and the high-precision wavelength calibration that relates the rotation angle of the ELL14 stage to a continuous spectral coordinate. Collectively, these procedures define the hardware platform on which the content-adaptive scanner described in Supplementary Note~\ref{supp:adaptive} operates.

\subsection{Experimental Setup}

\begin{figure*}[!tt]
  \centering
  \includegraphics[width=1.0\linewidth]{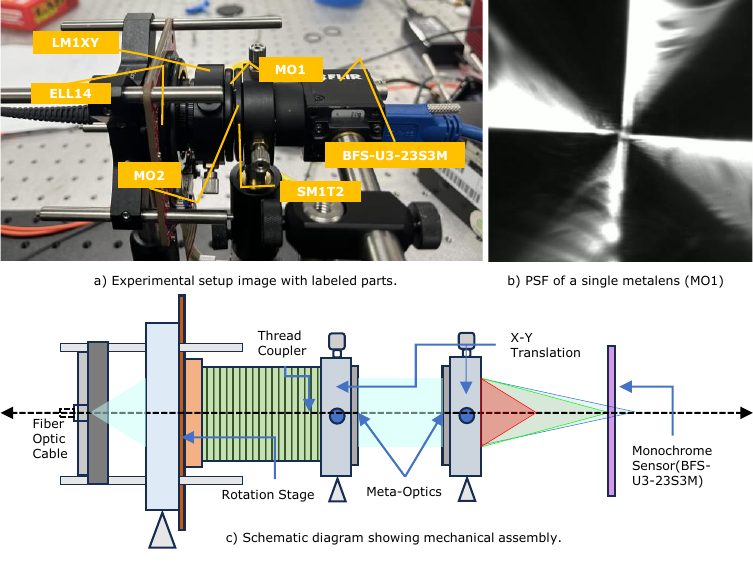}
  \caption{\textbf{Optomechanical assembly of the Moir\'e
meta-spectrometer.} \textbf{a,} Annotated image of the
prototype: the rotating meta-optic MO$_1$ is mounted on a two-axis
translation stage (LM1XY, Thorlabs) coupled to a piezo-motorised
rotation stage (ELL14) through SM1 thread adapters; the static
meta-optic MO$_2$ is mounted on its own LM1XY downstream, and the
focused PSF is recorded by a monochrome scientific camera
(BFS-U3-23S3M, FLIR Blackfly). \textbf{b,} Reference PSF through
MO$_1$ alone under quasi-incoherent monochromatic illumination,
used as the visual reference for the centring procedure of
Supplementary Note~\ref{supp:method:alignment}. \textbf{c,} Schematic of the optical train:
a fibre-coupled point-like source illuminates the cascaded
MO$_1$/MO$_2$ pair through SM1-threaded mounts; the wavelength
brought to focus on the sensor plane is selected by the relative
rotation angle $\theta$ commanded to the ELL14.}
  \label{fig:assembly}
\end{figure*}

The experimental setup consists of two meta-optics placed sequentially along a common optical axis. \textbf{$\text{MO}_1$} is the meta-optic mounted on a compact two-axis translation stage (LM1XY, Thorlabs), which is then joined to the ELL14 motorized rotation stage using external SM1 thread adapters. These adapters provide the required separation between components and avoid mechanical friction during rotation, while also keeping the optical axis mechanically well-defined through SM1-threaded interfaces. The X--Y translation stage provides a direct way to laterally shift $\text{MO}_1$ while the entire assembly still rotates, which is essential for centering the optic relative to the mechanical axis of the ELL14 piezo-motorized rotation stage. A dovetail translation is integrated at the base of the rotation-stage mount to decrease the separation distance between the two meta-optics. This axial degree of freedom allows us to bring $\text{MO}_1$ and $\text{MO}_2$ closer without disturbing the lateral centering, which is important for compact operation and for reducing propagation effects such as Talbot self-imaging over long inter-element distances.

$\text{MO}_2$ is the static meta-lens that is mounted only on an X--Y translation stage so that the centroid of $\text{MO}_2$ can coincide with the principal axis passing through the center of $\text{MO}_1$. In other words, $\text{MO}_1$ defines the rotating reference axis, and $\text{MO}_2$ is then translated to be co-axial with that reference. The $\text{MO}_2$ stage also uses a dovetail $z$-axis translation stage to reduce the distance between the meta-optics and to fine-tune the axial separation during alignment, especially when the PSF shape indicates residual defocus or spacing-induced changes in the effective system response.

The illumination source is placed approximately 40~mm to the left of $\text{MO}_{1}$ (approximately at a $2f$ distance) and consists of a circular slit of $50~\mu$m diameter coupled to a fiber probe. This configuration provides an approximately point-like input that is well-suited for monitoring the point spread function (PSF) of the combined system. We use a monochrome camera (FLIR Blackfly grayscale) as the detector, and we adjust the camera position and orientation such that (i) the PSF centroid is close to the center pixel and (ii) the dominant lobes are approximately equal in size and intensity. These conditions not only improve visual interpretability but also provide a practical symmetry check that the sensor plane is approximately perpendicular to the optical axis and that gross tilt is minimized.

\subsection{Alignment}
\label{supp:method:alignment}


The optical setup is extremely sensitive to lateral misalignment between the centroids of the two meta-optics. In this system, even small decenter between $\text{MO}_1$ and $\text{MO}_2$ can introduce angle-dependent aberrations and apparent motion of the focal spot, which directly degrades repeatability during rotational scanning. To ensure spectral and spatial stability of the focal point during rotation of $\text{MO}_1$, we use a PSF-driven visual centering procedure designed to satisfy three alignment objectives:
\begin{itemize}
    \item \textbf{Rotation-axis centering:} the axis of rotation of the motorized rotation stage coincides with the centroid of $\text{MO}_1$.
    \item \textbf{Co-axiality:} the centroids of $\text{MO}_1$ and $\text{MO}_2$ lie on the same optical axis.
    \item \textbf{Minimal tilt:} the surfaces of the two metalenses are parallel to each other and to the sensor plane, and approximately orthogonal to the line joining the point source and the sensor plane.
\end{itemize}
\textbf{Quasi-incoherent PSF as a stable alignment observable.} We illuminate the system with a monochromatic collimated laser module and make it quasi-incoherent by scrambling the phase with a rotating diffuser, which is a sandpapered translucent polymer sheet attached to a small motor. The reason for scrambling is that a purely coherent laser source generates strong speckle and interference artifacts at the detector, which makes the PSF fluctuate in time and biases any centroid-based centering. The rotating diffuser produces a time-varying phase screen that is averaged by the camera exposure, resulting in a stable PSF profile that is reproducible across repeated measurements. Under this illumination, the system PSF becomes an interpretable and repeatable visual reference that we monitor throughout alignment.

\subsection{Centering $\text{MO}_1$ to the rotation axis (ELL14).}
\begin{figure*}[!tt]
  \centering
  \includegraphics[width=1.0\linewidth]{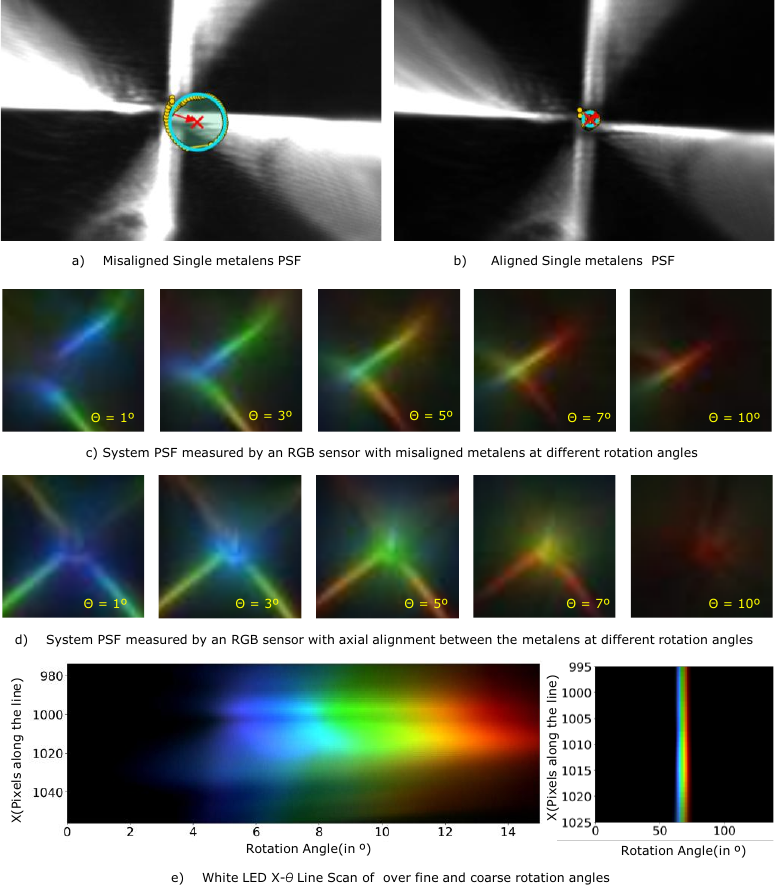}
  \caption{\textbf{PSF-driven alignment of the Moir\'e meta-optics.}
\textbf{a, b,} Single-metalens PSF through MO$_1$ before (\textbf{a}) and after (\textbf{b}) centering MO$_1$ on the ELL14 rotation axis: we track the PSF centroid over a full rotation (KLT tracker), fit a circle to its path, and translate MO$_1$ (LM1XY) until the fitted radius is below one binned pixel. \textbf{c, d,} Cascaded MO$_1$/MO$_2$ PSF (RGB sensor) at $\theta = 1^\circ, 3^\circ, 5^\circ, 7^\circ, 10^\circ$ before (\textbf{c}) and after (\textbf{d}) co-axially aligning MO$_2$ to MO$_1$: misalignment causes angle-dependent focal drift and asymmetric, fan-shaped lobes; alignment yields a stationary centroid and clean, wavelength-selective intensity modulation. \textbf{e,} $X$--$\theta$ line-scan of a white-LED source through the aligned system over fine ($\theta\in[0,14]^\circ$, left) and coarse ($\theta\in[0,100]^\circ$, right) ranges; the dominant bright track follows the calibrated $\lambda$--$\theta$ dispersion.}
  \label{fig:misalignment}
\end{figure*}
The first objective is to align the centroid of $\text{MO}_1$ with the axis of rotation of the motorized rotation stage. A lateral decenter from the rotation axis produces a characteristic and intuitive signature: during smooth rotation, the PSF centroid does not remain stationary, but instead traces a circular trajectory around the true rotation center in the detector plane. Conceptually, $\text{MO}_1$ rotates about a mechanical axis; if the optic center is offset from this axis, the optical response ``orbits'' around the true rotation center, and the PSF centroid draws an arc (or a full circle over a complete rotation). This trajectory provides a direct, image-domain indicator of decenter, and it is the basis of our centering procedure.

To quantify this trajectory and reduce observer bias, we use a KLT (Kanade-Lucas-Tomasi) feature tracker to track a stable, high-contrast region of the PSF (typically the bright core or a consistent portion of the fan-like PSF) across frames acquired during a rotation sweep. The tracked points form a 2D trajectory in detector coordinates. We then fit a circle to this trajectory and estimate the rotation center(PSF centroid) from the circle fit. We then use the LM1XY translation mount beneath $\text{MO}_1$ to apply small lateral corrections that move $\text{MO}_1$ toward this estimated center. After each adjustment, we repeat the rotation sweep and tracking. The procedure converges when the fitted radius becomes minimal and, visually, the PSF centroid remains nearly stationary over rotation, indicating that the centroid of $\text{MO}_1$ is well registered to the ELL14 rotation axis.

\textbf{Co-aligning $\text{MO}_2$ to $\text{MO}_1$ and stabilizing the focus during rotation.}
Once $\text{MO}_1$ is centered, we align $\text{MO}_2$ using its XY translation stage such that the combined system produces a stable focus whose position does not drift during rotation of $\text{MO}_1$. This step enforces co-axiality between the two meta-optics: $\text{MO}_1$ defines the rotating optical axis, and $\text{MO}_2$ is translated until the PSF symmetry and centroid stability are maximized. The main visual cue is that the $\text{MO}_1$--$\text{MO}_2$ pair behaves as an effective convex lens whose focusing behavior is controlled by the relative angle between the two meta-optics. Therefore, as $\text{MO}_1$ rotates about its axis, the image of the point source should remain spatially static on the sensor if $\text{MO}_2$ is properly centered and the two surfaces are approximately parallel. At the same time, the PSF intensity is expected to vary smoothly with angle, with a maximum at the relative angle corresponding to the monochromatic illumination wavelength (consistent with the wavelength-selective response of the meta-optic pair). In our experiments, successful alignment is indicated by (i) negligible lateral drift of the PSF centroid over the full rotation sweep, (ii) smooth and repeatable intensity modulation versus angle, and (iii) stable PSF morphology (no abrupt asymmetry changes) across repeated sweeps.

Finally, after co-alignment, we reduce the $MO_{1}-MO_{2}$ separation using the dovetail $z$-translations on the $\text{MO}_1$ and $\text{MO}_2$ mounts. This step is performed after lateral centering because changing axial spacing can slightly perturb mechanical stresses and alignment. We iteratively reduce the spacing while confirming that the PSF remains stable and centered, thereby achieving a compact configuration with reduced Talbot-length-related sensitivity while preserving repeatability.
\subsection{Precision calibration of the Moir\'e spectrometer}

Wavelength calibration of the spectrometer was performed by acquiring the angular emission spectra of six distinct laser configurations (407.08 nm, 533.09 nm, 636.91 nm, 787.47 nm, 849.72 nm, and 974.16 nm). To mitigate the effects of mechanical backlash and hysteresis in the motorized rotational stage, bidirectional spectral acquisitions—consisting of both clockwise and counterclockwise sweeps—were performed for each reference wavelength. The acquired intensity profiles from both scanning directions were averaged to yield a robust, directionally unbiased signal. Because the discrete angular sampling inherently limits the precision of raw peak identification (i.e., using a simple maximum), we determined the exact central angle of each emission line by fitting a Gaussian distribution to the averaged intensity profile surrounding the region of maximum intensity. The resulting highly precise Gaussian-fitted peak centers ($\theta_{peak}$) were mapped against the theoretical wavenumber ($10^7 / \lambda$, in cm$^{-1}$) of each corresponding laser. Linear regression analysis of these six points revealed a highly consistent calibration function, consistent with the linear wavenumber--angle dispersion of Eq.~\eqref{eq:calibration}: $\nu~[\mathrm{cm}^{-1}] = 358.4601\,\theta_{\mathrm{peak}} + 11761.6953$, with a coefficient of determination ($R^2$) exceeding 0.999. This experimentally derived linear map was subsequently rigorously inverted to establish the mapping from raw rotational angle to continuous wavelength ($\lambda = 10^7 / (358.4601 \times \theta_{peak} + 11761.6953)$), forming the definitive spectral calibration used across all subsequent commercial light source benchmarks and adaptive sampling routines. 




\bibliography{sn-bibliography}